\documentclass[aps,prl,twocolumn,showpacs]{revtex4}

\usepackage{graphicx}
\usepackage{amssymb}
\usepackage{amsfonts}
\usepackage{amsmath}
\usepackage{dcolumn}

\begin{document}
\title{Lifshitz Transitions in 122-Pnictides Under Pressure}

\author{Khandker Quader}
\affiliation{Department of Physics, Kent State University, Kent, OH 44242}
\author{Michael Widom}
\affiliation{Department of Physics, Carnegie-Mellon University, Pittsburgh, PA 15213}

\date{\today}

\begin{abstract}
We demonstrate, using $T=0$ first principles total energy calculations, that observed 
pressure-driven anomalies in the entire 122-pnictides
family ($A$Fe$_2$As$_2$; $A$ = alkali earth element Ca, Sr, Ba) can be understood as consequences 
of Lifshitz transitions (LTs)~\cite{Lifshitz60}. 
Our results for energy band dispersions and spectra, 
lattice parameters, enthalpies, magnetism, and elastic constants over a wide
range of hydrostatic pressure provide a coherent understanding of multiple transitions
in these compounds, namely, enthalpic,  magnetic and tetragonal (T) - collapsed tetragonal (cT) transitions. 
In particular, the T-cT transition and anomalies in lattice parameters and elastic properties, 
observed at finite temperatures, are interpreted as arising from 
proximity to $T=0$ Lifshitz transitions, wherein pressure causes non-trivial changes 
in the Fermi surface topology in these materials. 
\end{abstract} 
  

\maketitle

Lifshitz transitions (LTs)~\cite{Lifshitz60},
topological changes of a material's Fermi surface caused
by external pressure or chemical substitution, are of considerable current interest. 
Changes in electronic band structure, such as disruption or creation of a Fermi surface neck, and creation or disappearance of a pocket, constitute typical topological changes.
Because they require band dispersion extrema, LTs often appear on points or lines of high symmetry in the Brillouin zone. 
At zero temperature, $T= 0$, LTs are true phase transitions of order 2~1/2 (as in Ehrenfest's classification),
while at finite $T$, thermal smearing of the Fermi surface causes rapidly varying but analytic crossovers of properties~\cite{Varlamov89}. Resulting anomalies in lattice parameters, density of states near the Fermi energy
$E_F$, elastic properties, and electron dynamics manifest in observable
thermodynamic and transport properties~\cite{Lifshitz60,Varlamov89}.  Present-day angle resolved photoemission spectroscopy (ARPES) experiments are capable of mapping Fermi surface topology, and thus can provide a more direct signature of LTs.

The 122 pnictides, $A$Fe$_2$As$_2$ ($A$ = Ca, Sr, Ba), display structural, magnetic or superconducting phase
transitions upon doping or applied pressure\cite{Alireza09,Torikach09,Mani09,Saha11};
notably, a tetragonal phase (T) with a large $c$-axis, ``collapses'' to a phase (cT) with a smaller $c$-axis.
Both tetragonal phases share the space group I4/mmm.
In this paper based on $T = 0$ first principles total energy density functional theory (DFT) calculations as a function of pressure, we propose that the T-cT transition observed at finite temperature is actually a crossover behavior resulting from a $T=0$  Lifshitz transition. 
While at present there are no claims of observation of LTs in the pnictides under pressure, recent ARPES experiments~\cite{Liu10,LIu11} and theoretical interpretation in electron-doped Ba-122 provide evidence for LTs.
 Our results for energy band dispersions and spectra, $c$- and $a$-axis lattice parameters, magnetism, and elastic constants over a wide range of hydrostatic pressure, $P$ = 0 to 60 GPa, enable us to elucidate several additional pressure-driven features, and provide a more complete understanding of the 122-pnictides under pressure. 
We include in our results, $A$ = Ra, as it clarifies trends in behavior of the 122 series. 
 
In the past, theoretical evidence for LTs under pressure have been found, for example, in DFT calculations of Zn and Cd~\cite{Novikov99}, and Os~\cite{Koudela06}.
Effects of LTs on thermodynamic and transport properties may be small and hence pose challenges for experimentalists. Nevertheless,
experimental signatures of LTs have been deduced, for example, from pressure study of elastic constants in YCo$_5$~\cite{Rosner06}, 
Shubnikov-dHvA effect in Cd~\cite{Budko84}, and through the superconducting transition temperature (T$_c$) and magnetothermopower in Al~\cite{Overcash81}. 

The 122 pnictides share common structures (Fig. 1).
At ambient pressure, Ca-, Sr-, Ba-122 compounds exhibit transitions from the high-temperature tetragonal (T) phase
to a low temperature orthorhombic (OR) phase (space group Fmmm), striped along the $a$-axis and antiferromagnetically (AFM) ordered along the $c$-axis, at T$_{OR}$ $\sim$ 170 K, 205 K, 140 K respectively~\cite{Torikach08,Krellner08,Huang08}. 
This may be viewed as a magneto-structural transition from a high-T phase with fluctuating magnetic moments~\cite{Diallo10} to one with long-range AFM order.
The T-OR transition temperature $T_{OR}$ decreases with applied pressure.

\begin{figure}
\includegraphics[trim = 0in 6.5in 0in 0in, clip,width=3in]{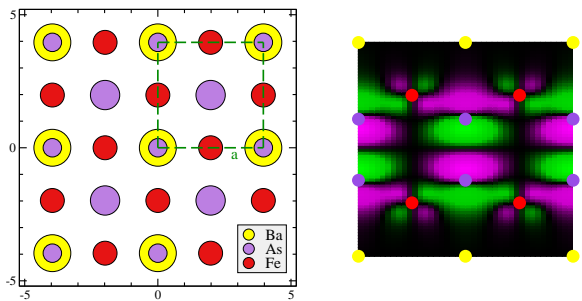}
\caption{Lattice Structure of 122 pnictides. The 122 pnictides $A$Fe$_2$As$_2$ 
($A$ = Ca, Sr, Ba) possess the same lattice structures. (left) Structure of Ba-122 tetragonal (T) phase viewed along the (001) axis, from $z=0$ to $c/2$. $xy$ coordinates are given in units of~\AA.  Atom size indicates $z$ position, with large below small. (right) Structure within the $xz$-plane at $y=0$. Atoms are color-coded as on left. Background illustrates a purely real M-point wavefunction (see text), with green indicating positive, and magenta negative values.}
\label{fig:struct}
\end{figure}

Under hydrostatic pressure, at low-$T$, the 122-pnictides lose their OR-AFM state.
Ca-122 and Sr-122 exhibit 1st-order transitions to the collapsed tetragonal (cT) phase with decreased $c$-axis value, at $P\sim$ 0.35 GPa~\cite{Canfield09},
and uncollapsed tetragonal (T) phase at $P\sim$ 4.4 GPa~\cite{ikeda12,kotegawa09} respectively.
Ba-122 appears to undergo a continuous (or weakly 1st-order) transition at $P\sim$ 10.5 GPa~\cite{Mittal11,Duncan10,Yamazaki10} to a T phase;
experiments are variously at T $\sim$ 33K and 100K. The critical pressures tend to be lower under non-hydrostatic conditions.
With increased pressure, the T phase $a$- and $c$-axis parameters decrease smoothly until pressures, 
$P\sim$ 7-10 GPa~\cite{ikeda12,kasinathan11,Uhoya11} and 
29 GPa~\cite{Mittal11,Uhoya10} respectively for Sr-122 and Ba-122, at which they evolve anomalously until the cT state is reached. High pressure x-ray diffraction studies of the Ba-122 and Sr-122 compounds attributed~\cite{Uhoya10,Uhoya11} these lattice parameter anomalies to a negative compressibility along the $a$-axis.

Previous DFT calculations~\cite{Mazin10,Tomic12,Colonna11} considered the pressure and doping dependence of Ba-122. Two pressure-driven transitions were
obtained~\cite{Tomic12,Colonna11}, and abrupt jumps in the bulk modulus reported~\cite{Tomic12} at both low and high pressure transitions. A DFT-based molecular dynamics (MD) calculation~\cite{Backes13} obtained two pressure-driven transitions in Ba-122 at low-$T$.  At higher $T$,  the sharp OR-T transition at $P\sim$ 12.5 GPa is shifted up to
15 GPa, and somewhat smoothed out, and at higher pressure the T-cT transition becomes almost indiscernible. 

Our DFT total energy study systematically explores the different pressure-driven transitions.  We begin with the T and OR-AFM states at $P=0$ and steadily increment the pressure, relaxing the structure at each $P$, and recording the equilibrium lattice parameters, total energy and magnetic moment.  To explore dependence on the alkali earth element $A$, in addition to Ca-, Sr- and Ba-122 pnictides, we also performed DFT calculations on RaFe$_2$As$_2$ and MgFe$_2$As$_2$ and compared their enthalpies to the pure elements and known stable binary phases. 
MgFe$_2$As$_2$ is found to be energetically unfavorable, both in its tI10 and oF20 forms. However, we predict that RaFe$_2$As$_2$, which may not be feasible to synthesize, is stable, and its pressure-driven transitions closely resemble those of Ba-122. 
We include this in our results as it clarifies trends in behavior of the 122 series.

Our calculation utilizes the plane-wave based DFT code VASP~\cite{Kresse96} with the all-electron projector augmented wave method~\cite{Kresse99} carried out in the Perdew-Burke-Ernzerhof generalized gradient approximation~\cite{PBE} to the exchange-correlation potential. Energy cutoffs and $k$-point meshes are increased to converge total energies to better than 1 meV/atom.  Fermi surface smearing of 0.2 eV is employed for relaxations and 0.05 eV for $\Delta H$ values. Precise transition pressures were found to be sensitive to smearing and energy cutoff. Densities of states utilize the tetrahedron method with subsequent Gaussian smearing of 0.01 eV. Calculations are performed within the unit cells with Pearson type tI10 for the tetragonal phase and oF20 for the orthorhombic phase. However, a $\sqrt{2}\times\sqrt{2}$ supercell of tI10 is employed for consistency with oF20 for calculations of $\Delta H$. Striped antiferromagnetic collinear spin configurations are utilized in the OR-AFM state.  To observe pressure-driven transitions we begin with $P=0$ structures and increase the pressure in steps of 1 GPa or less, fully relaxing the lattice parameters and internal coordinates at each step. Elastic constants were calculated within VASP by finite differences of stress with respect to strain.

\begin{table}
\begin{tabular}{r|rrr}
$A$ & $P_H$ & $P_M$ & $P_L$ \\
\hline
Ca  &  0.46 &   3.2   & -3.2 \\
Sr  &   6.7 &  9.3   &  7.5 \\
Ba  &  13.3 &  14.6   &   32 \\
Ra  &  19.2 &  20.4   &   51 \\
\end{tabular}
\caption{\label{tab:P} Calculated critical pressures for $T=0$ transitions common to the 122 pnictides. All pressures are in GPa. $P_H$  denotes enthalpic transition, in which the OR-AFM state loses thermodynamic stability to a tetragonal state. $P_M$ is the magnetic transition at which the OR-AFM state loses its magnetic moment ($M=0$). $P_L$ is identified as a Lifshitz transition (see text).}
 \end{table}

Our key results on the types of pressure-driven $T$=0 transitions are summarized in Table~\ref{tab:P} and Fig. 2. 
We find three types of transition to occur: an enthalpic transition, in which the OR-AFM state loses thermodynamic stability to a tetragonal state; a magnetic transition, in which the OR-AFM state loses its magnetic moment; a Lifshitz transition, in which the tetragonal T state collapses to cT. These transitions occur in different sequences depending on  $A$; hence some occur within metastable states and will not be observed
in experiments in thermodynamic equilibrium. 

\begin{figure}
\includegraphics[clip,angle=0,width=3in]{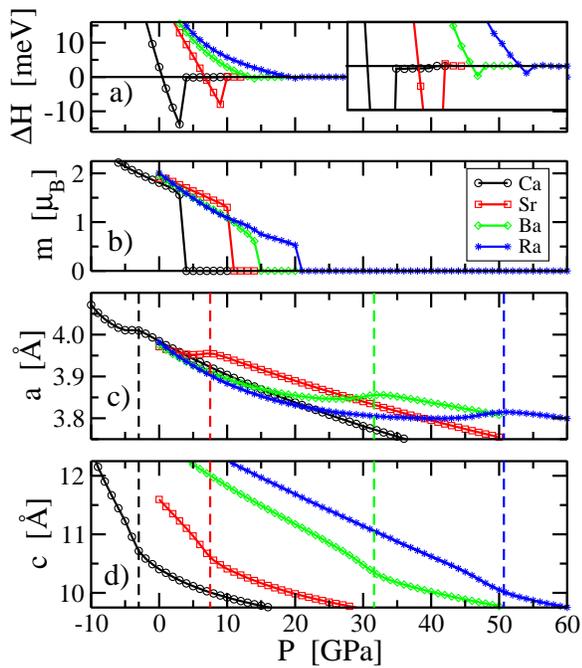}
\caption {Key pressure-dependent results for $A$-122 pnictides, $A$=Ca, Sr, Ba and Ra.
(a) Enthalpy differences $\Delta H=H_{\rm cT}-H_{\rm OR}$ between tetragonal  and OR-AFM phases; the inset enlarges the $\Delta H=0$ crossings. (b) Magnetic moments vanishing abruptly at the magnetic transition $P_M$. (c) and d) show the anomalous behavior of the  tetragonal $a$ and $c$ axis parameters.  Vertical dashed lines indicate Lifshitz transition pressures, $P_L$.}
\label{fig:allvsP}
\end{figure}

{\it Enthalpic transition at $P_H$} (Fig. 2a) :   For  $A$=Ca, both  OR-AFM and nonmagnetic states exist simultaneously at pressures above and below $P_H$. However, high-$P$ favors the lower volume cT state causing the enthalpy difference $\Delta H\equiv H_{\rm cT}-H_{\rm OR} = \Delta E - P \Delta V$ to change sign at $P_H$~\cite{Widom-Quader13}.  The situation is similar for $A$=Sr, Ba, Ra, with the exception that the transition from OR-AFM is to an uncollapsed tetragonal state T, rather than the collapsed cT. For all $A$, the OR-AFM state continues as a metastable state for $P>P_H$.  $P_H$ is progressively larger for $A$= Ca, Sr, Ba, Ra  because $\Delta E$ is greater, and the volume difference $\Delta V$ is smaller.  Because $\Delta H$ crosses zero linearly at $P_H$, these transitions are first order for all four 122 pnictides considered here.

{\it Magnetic transition at $P_M$} (Fig. 2b): For all four 122 compounds, the magnetic moments vanish suddenly, i.e. no stable OR-AFM state exists for $P>P_M$, even though it exists and maintains a large magnetic moment for $P_H<P<P_M$. We define $\Delta H=0$ for $P>P_M$ because OR-AFM structures relax to the nonmagnetic T (or cT) state.

{\it Lifshitz transition at $P_L$}: This separates the two tetragonal states, T and cT.  Fig. 2(c,d) shows that $P_L$ 
is characterized by anomalies in tetragonal $a$ and $c$ axis lattice parameters. 
The a axis varies non-monotonically, although the unit cell volume $V = a^2c$ decreases monotonically with $P$ as required by thermodynamics.
$P_L$ increases monotonically with $A$ as it advances down the periodic table. $P_L$ is negative for $A$=Ca, hence the transition at $P_H$ goes directly from OR to cT.
The transitions are fully reversible, as the lattice parameter curves exactly reproduce without hysteresis under increasing or decreasing pressure.

\begin{figure}
\includegraphics[clip,angle=0,width=3in]{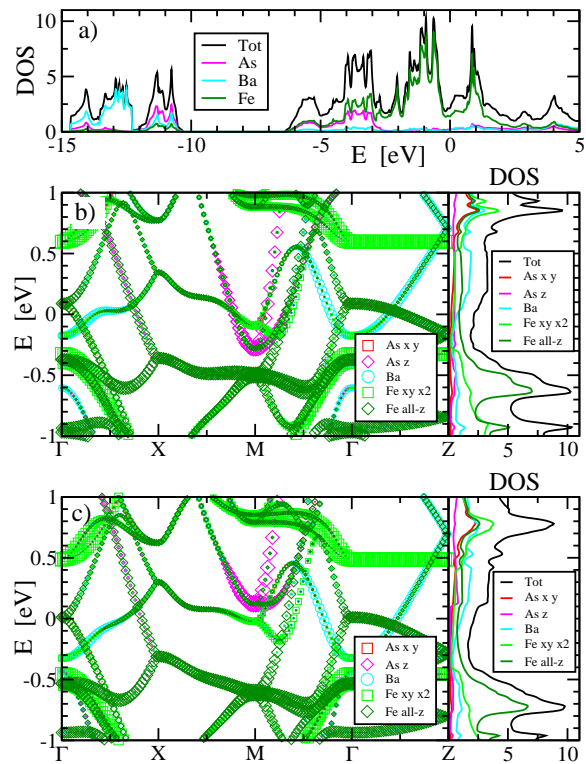}
\caption{Band structure and density of states (DOS) for Ba-122. (a) Total density of states (DOS) and Fe, As, and Ba partial DOS over a wide range of energy at $P=28$ GPa. 
(b) and (c) Band structures in the primitive tetragonal cell Brillouin zone at $P=28$ and $P=34$ GPa, i.e., below and 
above $P_L$ respectively. The corresponding pseudogap regions of total, and Fe, As, and Ba partial DOS around $E_F$ are shown alongside; the As and Ba DOS have been scaled 3x to make them more visible.
Note that the As $p_z$ electron pocket at the $M$ point moves above $E_F$ for $P>P_L$. Plotting symbol size indicates the projection onto atomic orbitals.} 
\label{fig:dos}
\end{figure}

Our electronic structure calculations give {\it band dispersions} along different k-directions, and also the density of states (DOS) integrated over the full Brillouin zone. Since the results are similar overall for all 122-pnictides, and some results for Ca-122 are presented elsewhere~\cite{Widom-Quader13}, 
we present in Fig. 3 dispersion relations for individual bands for Ba-122 at representative pressures as a prototypical case.
We draw particular attention to the band at the M-point, comprised mostly of anti-bonding As-$p_z$, mixed with bonding Fe-$d_{xz,yz}$ orbitals (see wavefunction in Fig. 1). As Fig. 3(b) shows, this band
lies below $E_F$ at $P=28$ GPa, forming an electron pocket, 
while Fig. 3(c) shows that at $P=34$ GPa, this band has moved to above $E_F$, so that the electron pocket has emptied; the transition occurs at $P_L$ = 31.6 GPa. 

Pressure dependence of this and other bands are illustrated Fig. 4(a,b).  This topological change in the Fermi surface correlates strongly with the anomalous behavior of the $a$ and $c$ lattice parameters discussed above and shown in Fig. 2. We take this band crossing to define the value of $P_L$, which we recognize as a Lifshitz transition.  While in principle this corresponds to a van Hove singularity in DOS that moves across $E_F$, this is not easy to discern among several DOS features characteristic of these materials.  Additional bands can be seen crossing $E_F$ in Fig. 3(a,b) indicating that many Lifshitz transitions exist, which is not surprising given the large number of bands in a crystal structure with many atoms per unit cell.

We point out an unusual feature at the $\Gamma$-point: A set of degenerate bands, dominated by Fe-$d_{xz,yz}$ orbitals, move from above $E_F$  for $P=28$ GPa (Fig. 3b) to a value very close to $E_F$ for $P=34$ GPa (Fig. 3c). These bands remain almost 
constant in energy for all $P > P_L$. Thus a hole pocket appears to shrink nearly to a point for $P > P_L$.  We find a similar degenerate band-pinning for Ra-122 (not shown) slightly below $E_F$, converting a hole pocket into a very small electron pocket. In the unfolded Brillouin zone this degeneracy lies at the M-point.

\begin{figure}
\includegraphics[width=3in,angle=0]{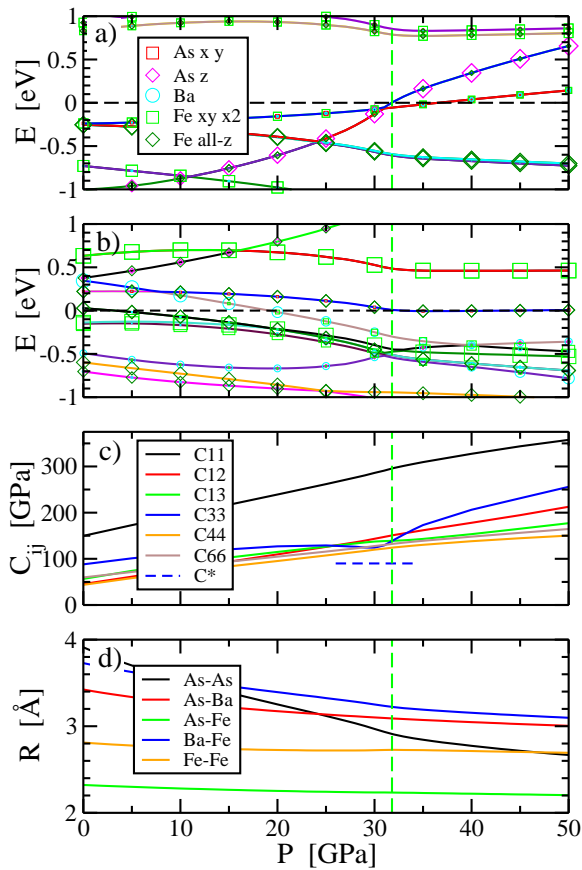}
\caption{Pressure variations of key quantities for Ba-122. (a) and (b) M- and $\Gamma$-point bands respectively (see text). Dashed vertical line locates $P_L$. (c) Tetragonal elastic constants calculated within VASP (see text).
The dashed horizontal line indicates the limit for elastic stability. (d) Interatomic separations --  various calculated bond-lengths with pressure (see text).}
\label{fig:BavsP}
\end{figure}

Fig. 4(c) shows the variation of the calculated tetragonal elastic constants with pressure.   
While most elastic constants increase monotonically with pressure, $C_{33}$, related to stress-strain along the $c$-axis, displays anomalous behavior around the T-cT Lifshitz transition pressure, $P_L$. Slightly below $P_L$, $C_{33}$ decreases with pressure to a minimum, and then increases again beyond $P_L$. As can be seen, $C_{33}$ lies above the limit of elastic stability,
i.e. $C_{33}>C^*=2 C_{13}^2/(C_{11}+C_{12})$=86 GPa, given the values of $C_{ij}$ at $P_L$.
Evidently, the T phase is heading towards an elastic instability in the vicinity of $P_L$ that is avoided by the transition into the cT state. We note that $C_{33}$, and hence compressibility, does not go negative, contrary to some views in literature~\cite{Uhoya10,Uhoya11}.

Fig. 4(d) 
shows the behavior of various calculated bond-lengths with pressure. The As-As bond length drops below 3~\AA, reaching 2.9~\AA~ as $P \rightarrow P_L$. As we discussed above, the M-point band that crosses $E_F$ at $P_L$ is dominated by As-$p_z$ orbitals between As atoms that are neighbors in the z-direction, and hence their separation closely tracks the variation of the $c$ axis.  As the As-As bond length drops, causing the energy of this repulsive anti-bond to rise towards $E_F$, the resulting depletion of the repulsive bond softens $C_{33}$. Following the collapse, other repulsive forces stabilize the structure with a reduced $c$-axis, as can be seen in the subsequent rise of $C_{33}$. The same wavefunction is bonding between As and Fe, so its depletion can be associated with the increase in the $a$ lattice parameter.  Presumably the value of $P_L$ grows with increasing atomic radius of the alkali earth element because greater pressure is required to drive the As-As bond length below 3~\AA.

Doping Ca-122 with Co stabilizes the uncollapsed T phase at $P=0$~\cite{Ran12}. According to our model, and verified by explicit calculation, carrier doping raises $E_F$, hence populating the anti-bonding orbital that is vacant for undoped Ca-122 at $P=0$ and moving the LT to increasingly positive pressure, consistent with experimental observation~\cite{Gati12}.

Our calculations show that the 122 pnictides,
under pressure, exhibit several features which are universal within this family. 
One key result is that at high pressures all the 122 members exhibit Lifshitz transitions characterized by the vanishing of an electron pocket at the M-point, as it moves 
from below to above the Fermi energy.  This correlates well with anomalies 
in lattice parameters and elastic constants around the Lifshitz transition pressure, $P_L$. We suggest that the finite-temperature T-cT transitions, that occur in all the 122 compounds, are consequences of the $T=0$ Lifshitz transitions. This provides a novel explanation for the observed anomalies in lattice parameters and elastic constants, and one that is different from those existent in literature.  We also find the existence of additional Lifshitz transition at other pressures.
At lower pressures, $P_H$, the 122-pnictides all undergo first-order enthalpic transitions  from a AFM-OR phases to non-magnetic T or a cT phases.  In all cases, metastable magnetic OR phases persist up to higher pressures, $P_M$ when
magnetism is lost by first -order transitions.
Our predictions are consistent with a large number experimental observations by different groups.

We thank Karin Rabe, Di Xiao, Paul Canfield and Alan Goldman for useful discussions.  We acknowledge the hospitality of Aspen Center for Physics, where part of the manuscript was written.

\bibliography{lifshitz}

\end{document}